# Relaxing Constraints in Anonymous Multi Agent Path Finding for Large Agents

Stepan Dergachev [1,2][0000-0001-8858-2831] and Avdeev Dmitry[1][0009-0009-7833-7838]

[1] HSE University, Moscow, Russia
[2] FRC CSC RAS, Moscow, Russia

**Abstract.** The study addressed the problem of Anonymous Multi-Agent Pathfinding (AMAPF). Unlike the classical formulation, where the assignment of agents to goals is fixed, in the anonymous MAPF setting it is irrelevant which agent reaches specific goal, provided that all goals are occupied. Most existing multi-agent pathfinding algorithms rely on a discrete representation of the environment (e.g., square grids) and do not account for the sizes of agents. This limits their applicability in real-world scenarios, such as trajectory planning for mobile robots in warehouses. Conversely, methods operating in continuous space typically impose substantial restrictions on the input data, such as constraints on the distances between initial and goal positions or between start/goal positions and obstacles. In this work, we considered one of the AMAPF algorithms designed for continuous space, where agents are modeled as disks of equal size. The algorithm requires a strict minimum separation of 4 agent radii between any start/goal positions. Proposed a modification aimed at relaxing the constraints and reduce this limit from 4 to $2\sqrt{3}$. We theoretically demonstrated that the proposed enhancements preserve original theoretical properties, including the guarantee that all agents will eventually achieve their goals safely and without collisions.

## 1 Introduction

The multi-agent pathfinding (MAPF) is a challenging problem in the field of artificial intelligence that arises in numerous practical domains, including multi-robot systems, logistics, video game development and etc. In the classical formulation, agents are assigned to predetermined, fixed goals. However, in many real-world scenarios, such as automated warehouses, agents are interchangeable, and it is irrelevant which agent reaches a particular goal, as long as all goals are eventually achieved. This variant of the problem is commonly referred to as Anonymous Multi-agent pathfinding (AMAPF) [1].

Many algorithms for both classical MAPF and its anonymous variant are built upon discrete representation of the environment, such as grids, and typically ignore the sizes of agents [2–5]. These assumptions limit their applicability in real-world scenarios. On the other hand, existing AMAPF methods that operate directly in continuous spaces usually impose a number of strict input constraints to ensure correctness [6–8] such as minimum separation requirements between start and goal positions or between these positions and static obstacles. Such constraints further limit the applicability of these approaches in practice.

In this work, we focus on one of the algorithms[7] of so-called AMAPF for Large Agents (AMAPF-LA), that is, an approach designed for continuous environments while explicitly accounting for agent size. In this formulation, obstacles are modeled as polygons, whereas agents are represented as disks of unit radius. The algorithm has two main requirements: *(i)* the distance between the start/goal positions of any two agents must be at least 4, and *(ii)* the distance between any start/goal position and obstacles must be at least $\sqrt{5}$. When we first examined the paper, we noted that it uses a specific theorem to ensure correct operation. In this paper we relax the first condition from 4 to $2\sqrt{3}$. At the same time, we theoretically demonstrate that this modification preserves the fundamental properties of the original algorithm, including the guarantee that all agents reach assigned goals.

## 2  Related work

The MAPF problem formulations can be subdivided into several categories. The classic formulation of the problem is one where agents are assigned to specific goals, meaning it is crucial that a particular agent reaches a particular goal. Another important variant is the Anonymous Multi-Agent Path Finding (AMAPF) problem, where agents are not assigned to specific goals[9]. If the agents' dimensions are also taken into account, the problem is referred to as (Anonymous) Multi-Agent Path Finding for Large Agents.

Finding optimal solution for MAPF problem, where each agent is assigned to a specific goal, has been shown to be NP-hard [10]. However, much research has concentrated on developing either optimal [11] or bounded-suboptimal [12] algorithms. In addition, several algorithms without formal optimality guarantees have been proposed, including MAPP [13], Push and Swap [14], Bibox [15], and Push and Rotate [16].

In contrast to MAPF, the Anonymous MAPF problem under the makespan metric admits a polynomial-time optimal solution, using algorithms for finding maximum flow in a graph [10]. However, applying these algorithms directly to large-scale problems is often impractical. In [17], the authors introduce the suboptimal and efficient algorithm TSWAP.

Several AMAPF algorithms also operate in continuous spaces, but most of them neglect either the presence of obstacles [6, 18] or the physical size of agents[19]. Methods that do consider obstacles typically impose restrictive assumptions, such as requiring obstacles of special structure [8], enforcing minimum separation between start and goal locations [20], and constraining the distance between these positions and obstacles [7].

Finally, several works have explored learning-based methods for AMAPF (e.g., reinforcement learning and imitation learning approaches) [21–24]. While these methods often demonstrate promising empirical performance, they generally lack theoretical guarantees of safety and goal reachability.

## 3 Problem Statement

Consider a set $M = \{1,2,\ldots,m\}$ of agents in a common workspace $W \subseteq \mathbb{R}^2$. Each agent is represented by an open disk of unit radius, and its state is defined by the position of the disk's center.

We define a set of starting positions for the agents as $\mathcal{S} = \{s_1, \ldots, s_m\}$, where each $s_i \in \mathbb{R}^2$. Similarly, let $\mathcal{T} = \{\tau_1, \ldots, \tau_m\}$ be a set of goal points, with $\tau_i \in \mathbb{R}^2$.

The workspace contains traversable areas and obstacle regions. The subset of the workspace occupied by obstacles is denoted by $O \subset W$. The traversable area is denoted as $W_{free} = W \setminus O$. Furthermore, we define the set of collision-free positions, where agents do not collide with any obstacles, as $\mathcal{F}$

$$\mathcal{F} = \{x \in W_{free} \mid \mathcal{B}_1(p) \cap O = \emptyset\} \tag{1}$$

Where $\mathcal{B}_1(p)$ is an open unit disk. We consider $T$ to be the continuous time set, $T = R^+ \cup \{0\}$. A path is a continuous mapping over time, $\pi: T \to \mathcal{F}$. Furthermore, a pair of paths $\pi$ and $\pi'$ is said to be non-conflicting if the agents moving along these paths do not collide.

$$\nexists t \in T : \mathcal{B}_1(\pi(t)) \cap \mathcal{B}_1(\pi'(t)) \neq \emptyset \tag{2}$$

Thus, the **problem** is to find a set of paths for all agents, $\Pi = \{\pi_1, \ldots, \pi_m\}$, such that all paths are pairwise non-conflicting, each agent's path starts at its respective initial position $\pi_i(0) = s_i$, and ends at a unique goal position. $\exists t \in T : \bigcup_{i \in M} \pi_i(t) = \mathcal{T}$.

We consider a set of paths $\Pi = \{\pi_i\}_{i \in M}$ to be a solution to the problem. We define the length of a path $\pi$ as $|\pi|$. Thus, the cost of a solution is the sum of the lengths of the paths:

$$sum\_of\_costs(\Pi) = \sum_{i=1}^{m} |\pi_i| \tag{3}$$

In this research, we imposed no requirements that optimize the cost of the solution; however, solutions with a lower cost are more relevant.

## 4 Anonymous Multi-Agent Pathfinding for Large Agents

To provide a foundation for the proposed improvement, we first outline the base algorithm [7]. This description highlights its core principles and the limitations that arise from its assumptions regarding the input data. In the subsequent section, we introduce a modified version of the algorithm designed to mitigate these limitations. The suggested method relaxes the constraints imposed on the initial and goal configurations while preserving the theoretical properties of the original approach.

## 4.1 Background

The main idea of the algorithm consists in the sequential selection and movement of individual agents from their start to their goal positions with a guarantee of collision avoidance.

At the first step, the shortest paths are computed for each agent–goal pair, after which the optimal assignment of goals to agents is found to minimize the total path length. As a result, a current set of paths is formed: $\Gamma = \{\gamma_1, \ldots, \gamma_m\}$, where $\gamma_i: [0,1] \to \mathcal{F}, \gamma_i(0) = s_i, \bigcup_{i \in M} \gamma_i(1) = \tau_i$. Then, a single goal $\tau$ is identified, which does not block the movement of other agents along the paths from $\Gamma$. Next, a start position s is selected from which the goal $\tau$ can be reached without collisions with other agents stationed at their start positions. The agent corresponding to position s moves to the goal $\tau$, after which the area $\mathcal{B}_1(\tau)$ occupied by this agent is marked as an obstacle. The positions s and $\tau$ are removed from the sets of start and goal positions $\mathcal{S}$ and $\mathcal{T}$, respectively. The procedure is repeated until all agents reach their goals.

For the correct operation of the algorithm, a number of requirements regarding the start and goal positions must be satisfied:

1. The pairwise distance between any two start or goal positions is at least 4.

$$\forall v, v' \in \mathcal{S} \cup \mathcal{T}, \ \|v - v'\|_2 \geq 4 \tag{4}$$

2. The distance between any start or goal position and any obstacle is at least $\sqrt{5}$

$$\forall v \in \mathcal{S} \cup \mathcal{T} \text{ and } \forall x \in O, \ \|v - x\|_2 \geq \sqrt{5} \tag{5}$$

For a more rigorous treatment of the algorithm, we introduce several definitions and statements. The proofs for these can be found in the original work [7].

**Theorem 1.** *A goal $\tau_i \in \mathcal{T}$ is called **standalone** if, given the current set of paths $\Gamma$, it satisfies the following condition:*

$$\forall \gamma_j \in \Gamma, \ i \neq j, \ 1 \leq i, j \leq m, \ \forall x \in \gamma_j \quad \mathcal{B}_1(x) \cap \mathcal{B}_1(\tau_i) = \emptyset. \tag{6}$$

In other words, a standalone goal does not conflict with the vicinity of any other path and, consequently, does not obstruct the movement of the agents.

**Lemma 1.** *Let $v \in \mathcal{S} \cup \mathcal{T}, x \in \mathcal{F}$, and suppose $\mathcal{B}_1(v) \cap \mathcal{B}_1(x) \neq \emptyset$. Then the path $\gamma$ connecting v to x ($\gamma(0) = v, \gamma(1) = x$) is a line segment that lies entirely within $\mathcal{F}$.*

**Theorem 2.** *Consider a set of paths $\Gamma = \{\gamma_i\}_{i \in M}$ such that $\forall i \in M, \gamma_i(0) = s_i, \bigcup_{i \in M} \gamma_i(1) = \mathcal{T}$, and the value $sum\_of\_costs(\Gamma)$ is minimal among all such path sets. Then, for this set of paths, a standalone goal always exists.*

Let us consider a detailed step-by-step description of the algorithm.
1. For each agent $i$, we construct a set of shortest paths $\Gamma'_i = \{\gamma'_{i,j}\}$ from its start position $s_i$ to all goal positions $\tau_j$.
2. Using the sets $\Gamma'_i$, we find a set of paths $\Gamma = \{\gamma_i\}_{i \in M}$ such that $\forall i \in M, \gamma_i(0) = s_i, \bigcup_{i \in M} \gamma_i(1) = \mathcal{T}$, and the value of $sum\_of\_costs(\Gamma)$ is minimized. In this case, the Hungarian algorithm [25] can be used to find the optimal assignment of goals.

3. Let us find a standalone goal τ for the set of paths Γ. Theorem 1 states that such a goal always exists. Let j be the index of the agent that are assigned to the goal τ, i.e., $\gamma_j(1) = \tau$.
4. We define the last agent $i$ that blocks path $\gamma_j$, i.e.:

$$\exists\, t \in [0,1]\colon \mathcal{B}_1\big(\gamma_j(t)\big) \cap \mathcal{B}_1(s_i) \neq \emptyset\,\nexists\, k \in M, t' > t\colon \mathcal{B}_1\big(\gamma_j(t')\big) \cap \mathcal{B}_1(s_k) \neq \emptyset. \quad (7)$$

5. If this agent $i$ does not exist, we move agent $j$ by path $\gamma_j$ to goal τ, add the path to the total set Π, remove $s_j$ and τ from $\mathcal{S}$ and $\mathcal{T}$, and add $\mathcal{B}_1(\tau)$ to the set of obstacles $\big(W_{free} = W_{free} \setminus \mathcal{B}_1(\tau),\, \mathcal{F} = \mathcal{F} \setminus \mathcal{B}_2(\tau)\big)$.
6. Otherwise, agent $i$ is assigned a path constructed from the straight-line segment $[s_i, \gamma_j(t)]$ and the remaining part of the path from $\gamma_j(t)$ to τ. We move the agent along the new path, after which we add this path to Π, remove the start position $s_i$ and goal τ from their respective sets, and block the area $\mathcal{B}_1(\tau)$.
7. Repeat steps 1-6 until all agents are assigned goals.

Figure 1 illustrates one step of the algorithm's operation. Agents $i$ and $k$ (green disks) need to be assigned goals $t_i$ and $t_k$ (violet disks). An optimal set of paths was constructed. Goal $t_k$ was considered as a standalone goal; however, the path of agent k conflicted with the start position of agent $i$. Agent $i$ is assigned goal $t_k$, and a new path is constructed for it (red dashed line).

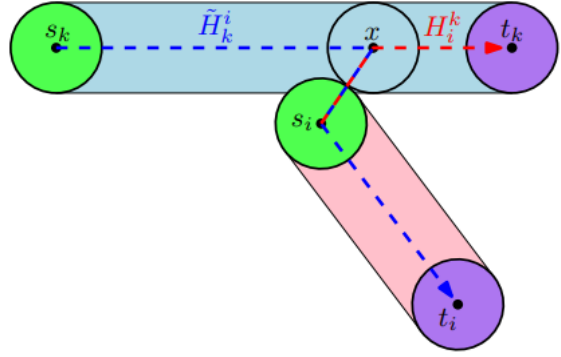

**Fig. 1.** Illustration of the basic conflict resolution method. The figure is taken from paper [7]

## 4.2 Suggested Method

This section will describe a modification of the algorithm previously considered from the paper [7]. For the subsequent discussion and proofs, we will leverage the previously introduced theorems and formulations. Let us describe the modified algorithm.

The modified algorithm proposes a different approach to handle the movement of agents when the theorems from the original algorithm in paper [7] are violated. When constraints are weakened, the conditions for these theorems may no longer hold. Consequently, a situation may occur where the distance between point $\gamma_j(t)$ and goal $\tau$ becomes less than 2, which could potentially lead to conflicts in the future. Therefore, in such cases, the algorithm selects a point located exactly at a distance of 2 from $\tau$ instead of $\gamma_j(t)$.

Let us consider a detailed step-by-step description of the modified algorithm.
1. For each agent $i$, we construct a set of shortest paths $\Gamma'_i = \{\gamma'_{i,j}\}$ from its start position $s_i$ to all goal positions $\tau_j$.
2. Using the sets $\Gamma'_i$, we find a set of paths $\Gamma = \{\gamma_i\}_{i \in M}$ such that $\forall i \in M$, $\gamma_i(0) = s_i$, $\bigcup_{i \in M} \gamma_i(1) = \mathcal{T}$, and the value of $sum\_of\_costs(\Gamma)$ is minimized. In this case, the Hungarian algorithm [25] can be used to find the optimal assignment of goals.
3. Let us find a standalone goal $\tau$ for the set of paths $\Gamma$. Theorem 1 states that such a goal always exists. Let j be the index of the agent that are assigned to the goal $\tau$, i.e., $\gamma_j(1) = \tau$.
4. We define the last agent $i$ that blocks path $\gamma_j$, i.e.:

$$\exists t \in [0,1]: \mathcal{B}_1\left(\gamma_j(t)\right) \cap \mathcal{B}_1(s_i) \neq \emptyset \nexists k \in M, \ t' > t: \mathcal{B}_1\left(\gamma_j(t')\right) \cap \mathcal{B}_1(s_k) \neq \emptyset. \quad (8)$$

5. If this agent $i$ does not exist, we move agent $j$ by path $\gamma_j$ to goal $\tau$, add the path to the total set $\Pi$, remove $s_j$ and $\tau$ from $\mathcal{S}$ and $\mathcal{T}$, and add $\mathcal{B}_1(\tau)$ to the set of obstacles $\left(W_{free} = W_{free} \setminus \mathcal{B}_1(\tau), \ \mathcal{F} = \mathcal{F} \setminus \mathcal{B}_2(\tau)\right)$.
6. Otherwise:
   a) If the distance to $\gamma_j(t)$ is at least 2, agent $i$ is assigned a path that consists of the straight-line segment $[s_i, \gamma_j(t)]$ and the remaining part of the path from $\gamma_j(t)$ to $\tau$. We move the agent along the new path, after which this path is added to $\Pi$, the start position $s_i$ and goal $\tau$ are removed from their respective sets, and the area $\mathcal{B}_1(\tau)$ is blocked (see Figure 2)
   b) If the distance to $\gamma_j(t)$ is less than 2, a single point $\gamma_j(t')$ at a distance of 2 from $\tau$ is chosen, i.e.:

$$\exists! \ t' \in [0,1] : \|\gamma_j(t'), \tau\|_2 = 2. \quad (9)$$

   Agent $i$ is assigned a path constructed from the straight-line segment $[s_i, \gamma_j(t')]$ and the remaining part of the path from $\gamma_j(t')$ to τ. We move the agent along the new path, after which we add this path to Π, remove the start position $s_i$ and goal τ from their respective sets, and block the area $\mathcal{B}_1(\tau)$ (see Figure 3).
7. Repeat steps 1-6 until all agents are assigned goals.

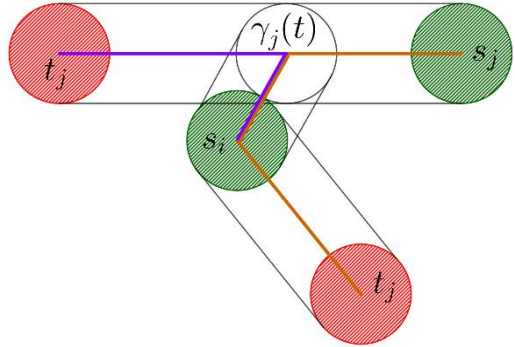

**Fig. 2.** Movement of agents in case of sufficient distance between $\gamma_j(t)$ and τ.

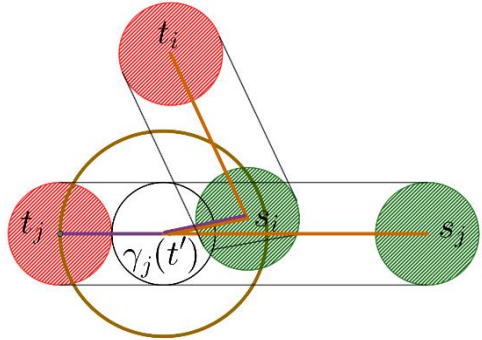

**Fig. 3.** Movement of agents in case of violation of the distance condition between $\gamma_j(t)$ and τ.

Next, we provide a proof of the algorithm's theoretical properties.

### 4.3 Theoretical Analysis

**Lemma 2.** Given a triangle ABC with $|AB| = 2$, $|AC| \geq 2$, $|BC| \geq 2\sqrt{3}$, prove that the distance from point C to the segment AB is at least 2.
*Proof.* Drop a perpendicular CH to the line AB. If $|CH| \geq 2$, then the distance to the segment is clearly at least 2, and the statement is proven. Now suppose $|CH| \leq 2$.

Assume that H lies on the segment AB. Then, by the Pythagorean theorem, $2\sqrt{3} \leq |CB| = \sqrt{|CH|^2 + |HB|^2} \leq \sqrt{|CH|^2 + |AB|^2} \leq \sqrt{2^2 + 2^2} = 2\sqrt{2}$, which is a contradiction. Hence, H does not lie on the segment AB, meaning the distance from C to the segment AB in this case is the distance to either point A or point B. By the given conditions, $|AC|, |BC| \geq 2$, which completes the proof.

**Theorem 3.** When executing step 6(a), the movement of the conflicting agent does not cause conflicts with other agents

*Proof.* When executing step 6(a), the agent moves a distance of 2 toward point $\gamma_j(t)$, and by assumption, the path starting from this point is conflict-free. Therefore, it remains only to verify that the movement over the distance of 2 is valid; the remainder of the path is valid by the definition of the chosen point. Note that there are no agents within a distance of 2 from point $\gamma_j(t)$, as otherwise this point would not be the last conflict point. Moreover, if there are no agents/goals within a distance of $2\sqrt{3}$ from the conflicting agent, then these two circles, inside which no agents are present, entirely cover the movement of the conflicting agent. Indeed, if these two circles did not cover the movement of the conflicting agent, then there would exist some point $w$ on the boundary of the union of these two circles such that the distance from w to the segment $[s_i, \gamma_j(t)]$ would be less than 2, i.e., $dist(w, [s_i, \gamma_j(t)]) < 2$. However, by assumption, the following conditions hold simultaneously: $dist(s_i, w) \geq 2\sqrt{3}, dist(\gamma_j(t), w) \geq 2$, and $dist(\gamma_j(t), s_i) = 2$. This leads to a geometry problem, the proof of which is given in Lemma 2. An illustration of the proof is shown in Fig. 4. The conflicting agent is marked as a green disk centered at B. The movement occurs from point B to point A. By definition, no agents can be inside the blue circles. Furthermore, during the movement from B to A, the distance from the center of the agent to the boundary of one of the two blue circles is always at least 2, as can be seen from the two nested circles of radius 2.

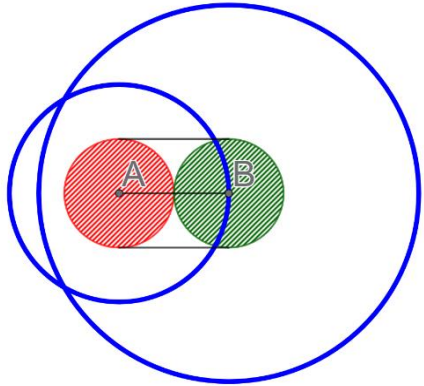

**Fig. 4.** Illustration of the proof of the correctness of movement along the new path

**Lemma 3**. *During the execution of the algorithm, $\forall j, 1 \leq j \leq m, \exists! t' \in [0,1]: dist(\gamma_j(t'), t_j) = 2$. Furthermore, $\gamma_j(t', 1) = [\gamma_j(t'), t_j]$. In other words, the point $\gamma_j(t')$ defined in step 6(b) is unique, and the path between $\gamma_j(t')$ and $t_j$ is a straight line segment.*

*Proof.* Consider the assignment $\Gamma'$ obtained during step 1. From Lemma 1 it follows that for any such point $\gamma_j(t')$ in step 6(b), the path between $\gamma_j(t')$ and $t_j$ is obstacle-free. Consequently, the algorithm will assign the optimal route as a straight line segment. Moreover, if there were multiple such points, the algorithm would have constructed a segment to the goal from the earliest such point, again contradicting the optimality of the path assignment.

**Lemma 4.** *Given a triangle ABC with $|AB| = 2, |BC| \leq 2, |AC| \geq 2\sqrt{3}$. Also given a point D such that $|CD| \geq 2\sqrt{3}, |AD| \geq 2\sqrt{3}$. Prove that $dist(BC, D) \geq 2$ and $dist(AB, D) \geq 2$.*

*Proof.* Let us prove that $dist(BC, D) \geq 2$. Drop a perpendicular $DH$ to the line $BC$. If $|DH| \geq 2$, then this point is proven. Now suppose $|DH| < 2$. Note that in this case $|CH| = \sqrt{|CD|^2 - |HD|^2} \geq 2\sqrt{2}$. In this case, $|CH| > |BC|$, meaning $H$ does not lie on $BC$, so the distance to the segment $BC$ is the distance to either point $B$ or $C$. Point $C$ already has a distance greater than 2 to point $D$ due to the constraint $2\sqrt{3}$. It remains to prove that $|BD| \geq 2|$ and then the problem is solved. Suppose this is not true. Then, due to the length constraints, angles $\angle ABC \geq \frac{2\pi}{3}, \angle CBD > \frac{2\pi}{3}, \angle ABD > \frac{2\pi}{3}$, by the Law of Cosines. But then it turns out that $\angle ABC + \angle CBD + \angle ABD > 2\pi$. If so, then point $B$ cannot be inside or on the boundary (possibly degenerate) of triangle $ACD$, since then the sum of angles would be $2\pi$. However, since all sides of triangle $ACD$ are greater than $2\sqrt{3}$ and $B$ is outside this triangle, then one of the points $A, C, D$ is at a distance greater than 2 from point $B$, which contradicts the constraints on the lengths of segments $AB, CB, DB$. Indeed, one of the sides must be greater than 2, because otherwise triangle $ACD$ would lie strictly inside a semicircle of radius 2, which is impossible, since the perimeter of the triangle is less than the perimeter of the semicircle $(4 + 2\pi)$, due to convexity and the nesting of convex figures, and since all sides are at least $2\sqrt{3}$, the perimeter of the triangle is at least $6\sqrt{3} > 4 + 2\pi$, a contradiction. Therefore, the distance to point $B$ in this case is also at least 2, so the problem for segment $BC$ is solved.

The case for segment AB is analogous.

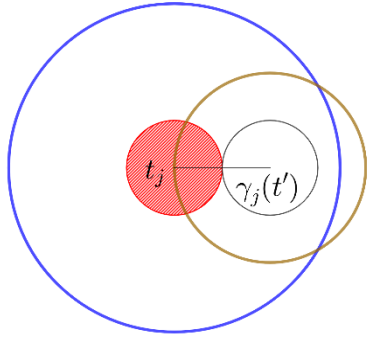

**Fig. 5.** The area in which the agent is located in case 6(b)

**Theorem 4.** *When executing step 6(b), the movement of the conflicting agent does not cause conflicts with other agents.*

*Proof.* First, note that when executing step 6(b), there exists a specific 'small' bounded region, which is formed by the intersection of the circle of radius 2 centered at $\gamma_j(t')$ and the set of all points that are at least $2\sqrt{3}$ away from $t_j$. An illustration of such region is provided in Figure 5.

It can be seen that the conflicting agent is always inside such region. If the conflicting agent were outside, step 6(a) would be executed instead. Indeed, if the conflicting agent $i$ is outside of this region, then $dist([\gamma_j(t'), t_j], s_i) > 2 \implies [\gamma_j(t'), t_j] \cap B_1(s_i) = \emptyset$, meaning the point $\gamma_j(t)$ defined in step 6(a) would be at least at distance 2 from the standalone goal $t_j$ by Lemma 3, so step 6(a) would be executed.

Therefore, only the case where the conflicting agent $i$ is inside such region remains. We now prove that the path from any inner point to the point $\gamma_j(t')$ from step 6(b) does not conflict with other agents. This requires solving a geometric problem, which is addressed in Lemma 4.

In the geometric interpretation, points $A, B,$ and $C$ correspond to $t_j, \gamma_j(t'),$ and $s_i$, respectively. Moreover, point $D$ represents some other agent. Accordingly, Lemma 4 shows that the polyline $CBA$ is at least at distance 2 from point $D$, which in our case implies that no conflicts occur during the movement. An illustration is provided in Fig.6.

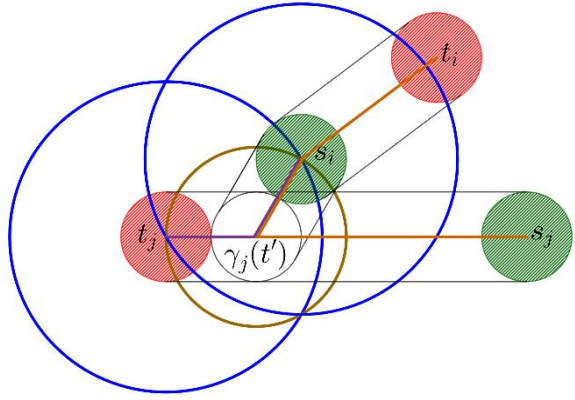

**Fig. 6.** Illustration of the proof of the correctness of movement along the new path

**Theorem 5.** *The standalone goal always exists.*

*Proof.* We base the beginning of the proof on the work from [7]. Note that if there is no standalone goal, then there exists a cycle of goals that conflict with each other. Without loss of generality, assume that for agent $i$, the start position is $s_i$ and the goal is $t_i$, and goal $t_{i+1}$ intersects the path of agent $i$ for each $i$ from 1 to $l - 1$ inclusive, and goal $t_1$ intersects the path of agent 1. Now we need to prove that there exists a set of paths assigning each agent $i$ to goal $t_{i+1}$ (for $i = l$, the goal is $t_1$), for $i$ from 1 to $l$, such that the total length of the new paths is better than the original. However, note that in the described modified algorithm, the conflicting agent moved less than 2 towards a point on the path that was at least 2 away from the standalone goal. We apply the algorithm to the path from $s_i$ to $t_i$, where the conflicting agent in this case is goal $t_{i+1}$. Applying all the above proofs to the modified algorithm, we conclude that the new assignment is strictly better than the old one, contradicting the initial assignment choice.

**Theorem 6.** *The algorithm is complete and finds a solution that is at most worse than the optimal solution by 4m radius of disk, provided the following conditions hold:*
1. *The pairwise distance between the centers of agents/goals is at least $2\sqrt{3}$*
2. *The distance between the centers of agents/goals and obstacles is at least $\sqrt{5}$*

*Proof.* To prove the algorithm's completeness, we must verify that during its execution, no collisions occur with obstacles, and no collisions occur between agents. Conflicts with obstacles are prevented by Lemma 1. Theorems 3 and 4 prove that when executing these steps, no conflicts between agents arise during movement. Therefore, no conflicts occur, meaning the algorithm always finds some solution. Moreover, when executing one of the sub-steps in step 6, the additional movement adds at most 4 to the cost, as noted in Theorems 3 and 4. Hence, the found solution is at most worse than the optimal solution by $4m$. Furthermore, after each algorithm iteration, $B_1(t_j)$ is added to the obstacles. Due to the $2\sqrt{3}$ constraint between agents and goals, the distance between agents and obstacles remains at least $\sqrt{5}$, since $\sqrt{5} < 2\sqrt{3}$. The $2\sqrt{3}$ constraint also persists, as the algorithm proceeds to a subset of the set for which this constraint held. Thus, the algorithm's constraints are preserved after each iteration.

## 5 Conclusion

In this work, we considered the problem of constructing a set of conflict-free paths for moving a group of agents from initial positions to a set of target positions, where it is not important which agent occupies which target. We examined an algorithm proposed in [7], which allows solutions to be found on a plane while accounting for the sizes of the agents. However, the original algorithm imposes constraints on the relative arrangement of the initial and target positions. In this work, we proposed a modification that relaxes these constraints by 13% while preserving the theoretical properties of the original algorithm. This improvement can help find solutions in problems with a higher density of agents. In cases where the original algorithm was inapplicable due to its constraints, the modified algorithm can successfully solve the problem.

An important direction for future research is to conduct experimental testing of the proposed method and compare it with existing alternatives, as well as to further relax the constraints on possible input data for multi-agent path planning algorithms.

8. Banyassady B, de Berg M, Bringmann K, et al (2022) Unlabeled Multi-Robot Motion Planning with Tighter Separation Bounds. In: 38th International Symposium on Computational Geometry (SoCG 2022)

9. Stern R, Sturtevant NR, Felner A, et al (2019) Multi-Agent Pathfinding: Definitions, Variants, and Benchmarks. In: Proceedings of the 12th Annual Symposium on Combinatorial Search (SoCS 2019). pp 151–158

10. Yu J, LaValle SM (2013) Multi-agent path planning and network flow. In: Algorithmic Foundations of Robotics X: Proceedings of the Tenth Workshop on the Algorithmic Foundations of Robotics. Springer, pp 157–173

11. Sharon G, Stern R, Felner A, Sturtevant NR (2015) Conflict-Based Search for Optimal Multi-Agent Pathfinding. Artificial Intelligence 219:40–66

12. Li J, Ruml W, Koenig S (2021) EECBS: A Bounded-Suboptimal Search for Multi-Agent Path Finding. In: Proceedings of the 35th AAAI Conference on Artificial Intelligence (AAAI 2021). pp 12353–12362

13. Wang K, Botea A (2011) MAPP: a scalable multi-agent path planning algorithm with tractability and completeness guarantees. Journal of Artificial Intelligence Research 42:55–90

14. Luna R, Bekris KE (2011) Push and swap: Fast cooperative path-finding with completeness guarantees. In: IJCAI. pp 294–300

15. Surynek P (2009) A novel approach to path planning for multiple robots in biconnected graphs. In: 2009 IEEE international conference on robotics and automation. IEEE, pp 3613–3619

16. De Wilde B, Ter Mors AW, Witteveen C (2014) Push and rotate: a complete multi-agent pathfinding algorithm. Journal of Artificial Intelligence Research 51:443–492

17. Okumura K, Défago X (2023) Solving simultaneous target assignment and path planning efficiently with time-independent execution. Artificial Intelligence 321:103946

18. Panagou D, Turpin M, Kumar V (2019) Decentralized Goal Assignment and Safe Trajectory Generation in Multirobot Networks via Multiple Lyapunov Functions. IEEE Transactions on Automatic Control 65:3365–3380

19. Adler A, De Berg M, Halperin D, Solovey K (2015) Efficient multi-robot motion planning for unlabeled discs in simple polygons. In: Algorithmic Foundations of Robotics XI: Selected Contributions of the Eleventh International Workshop on the Algorithmic Foundations of Robotics. Springer, pp 1–17

20. Turpin M, Mohta K, Michael N, Kumar V (2014) Goal Assignment and Trajectory Planning for Large Teams of Interchangeable Robots. Autonomous Robots 37:401–415

21. Lowe R, Wu YI, Tamar A, et al (2017) Multi-Agent Actor-Critic for Mixed Cooperative-Competitive Environments. Proceedings of the Advances in neural information processing systems ({NIPS} 2017) 30:

22. Ji X, Li H, Pan Z, et al (2021) Decentralized, Unlabeled Multi-Agent Navigation in Obstacle-Rich Environments Using Graph Neural Networks. In: Proceedings of IEEE/RSJ International Conference on Intelligent Robots and Systems (IROS 2021). pp 8936–8943